\documentclass[twoside]{IEEEtran}

\usepackage{cite}

\usepackage{amsmath,amssymb,amsfonts,mathrsfs,bm}

\usepackage{graphicx}
\usepackage{subfigure}
\graphicspath{{figures/}}

\usepackage{tabularx}
\usepackage{booktabs}

\makeatletter
  \newif\if@restonecol
\makeatother

\usepackage[linesnumbered,ruled,lined]{algorithm2e}
\usepackage{algpseudocode}

\usepackage{flushend}

\usepackage{scalerel}
\usepackage{tikz}
\usetikzlibrary{svg.path}

\definecolor{orcidlogocol}{HTML}{A6CE39}
\tikzset{
  orcidlogo/.pic={
    \fill[orcidlogocol] svg{M256,128c0,70.7-57.3,128-128,128C57.3,256,0,198.7,0,128C0,57.3,57.3,0,128,0C198.7,0,256,57.3,256,128z};
    \fill[white] svg{M86.3,186.2H70.9V79.1h15.4v48.4V186.2z}
                 svg{M108.9,79.1h41.6c39.6,0,57,28.3,57,53.6c0,27.5-21.5,53.6-56.8,53.6h-41.8V79.1z M124.3,172.4h24.5c34.9,0,42.9-26.5,42.9-39.7c0-21.5-13.7-39.7-43.7-39.7h-23.7V172.4z}
                 svg{M88.7,56.8c0,5.5-4.5,10.1-10.1,10.1c-5.6,0-10.1-4.6-10.1-10.1c0-5.6,4.5-10.1,10.1-10.1C84.2,46.7,88.7,51.3,88.7,56.8z};
  }
}

\newcommand\orcidicon[1]{%
  \href{https://orcid.org/#1}{
    \mbox{\scalerel*{
      \begin{tikzpicture}[yscale=-1, transform shape]
        \pic{orcidlogo};
      \end{tikzpicture}
    }{|}%
    }%
  }%
}

\usepackage[colorlinks,allcolors=black]{hyperref} 

\newtheorem{theorem}{Theorem}
\newtheorem{lemma}{Lemma}
\newtheorem{remark}{Remark}
\newtheorem{assumption}{Assumption}

\title{Adaptive Safety Evaluation for Connected and Automated Vehicles with Sparse Control Variates}

\author{Jingxuan Yang\textsuperscript{\orcidicon{0000-0001-9798-7347}}, Haowei Sun\textsuperscript{\orcidicon{0000-0002-0232-117X}}, Honglin He\textsuperscript{\orcidicon{0000-0003-4673-5283}}, Yi Zhang\textsuperscript{\orcidicon{0000-0001-5526-866X}}, \IEEEmembership{Member,~IEEE},\\ Shuo Feng\textsuperscript{\orcidicon{0000-0002-2117-4427}}, \IEEEmembership{Member,~IEEE} and Henry X. Liu\textsuperscript{\orcidicon{0000-0002-3685-9920}}, \IEEEmembership{Member,~IEEE}%
\thanks{This work is supported by National Key Research and Development Program under Grant 2021YFB2501200 and National Natural Science Foundation of China under Grant 62133002. \textit{(Corresponding author: Shuo Feng.)}}%
\thanks{Jingxuan Yang and Honglin He are with the Department of Automation, Tsinghua University, Beijing 100084, China (email: \{yangjx20, hehl21\}@mails.tsinghua.edu.cn).}%
\thanks{Haowei Sun and Henry X. Liu are with the Department of Civil and Environmental Engineering, University of Michigan, Ann Arbor, MI 48109, USA (e-mail: \{haoweis, henryliu\}@umich.edu).}%
\thanks{Yi Zhang is with the Department of Automation, Beijing National Research Center for Information Science and Technology (BNRist), Tsinghua University, Beijing 100084, China (e-mail: zhyi@tsinghua.edu.cn).}%
\thanks{Shuo Feng is with the Department of Automation, Tsinghua University, Beijing 100084, China and the University of Michigan Transportation Research Institute, Ann Arbor, MI 48109, USA (e-mail: fshuo@umich.edu).}
}

\begin{document}

\maketitle

\begin{abstract}
  Safety performance evaluation is critical for developing and deploying connected and automated vehicles (CAVs). One prevailing way is to design testing scenarios using prior knowledge of CAVs, test CAVs in these scenarios, and then evaluate their safety performances. However, significant differences between CAVs and prior knowledge could severely reduce the evaluation efficiency. Towards addressing this issue, most existing studies focus on the adaptive design of testing scenarios during the CAV testing process, but so far they cannot be applied to high-dimensional scenarios. In this paper, we focus on the adaptive safety performance evaluation by leveraging the testing results, after the CAV testing process. It can significantly improve the evaluation efficiency and be applied to high-dimensional scenarios. Specifically, instead of directly evaluating the unknown quantity (e.g., crash rates) of CAV safety performances, we evaluate the differences between the unknown quantity and known quantity (i.e., control variates). By leveraging the testing results, the control variates could be well designed and optimized such that the differences are close to zero, so the evaluation variance could be dramatically reduced for different CAVs. To handle the high-dimensional scenarios, we propose the sparse control variates method, where the control variates are designed only for the sparse and critical variables of scenarios. According to the number of critical variables in each scenario, the control variates are stratified into strata and optimized within each stratum using multiple linear regression techniques. We justify the proposed method's effectiveness by rigorous theoretical analysis and empirical study of high-dimensional overtaking scenarios. 
\end{abstract}

\begin{IEEEkeywords}
  Adaptive safety evaluation, connected and automated vehicles, sparse control variates, high-dimensional scenarios
\end{IEEEkeywords}

\section{Introduction}

\IEEEPARstart{T}{esting} and evaluation of safety performance are major challenges for the development and deployment of connected and automated vehicles (CAVs). One proposed way is to test CAVs in the naturalistic driving environments (NDE) through a combination of software simulation, test tracks, and public roads, observe their performances, and make statistical comparisons with human drivers. Due to the rarity of safety-critical events in NDE, however, hundreds of millions of miles and sometimes hundreds of billions of miles would be required to demonstrate CAVs' safety performance at the human-level \cite{kalra2016driving}, which is intolerably inefficient. To improve the efficiency and accelerate the evaluation process, the past few years have witnessed increasingly rapid advances in the field of testing scenario library generation (TSLG) \cite{li2021scegene,wang2021advsim,menzel2018scenarios,tian2018deeptest,rempe2022generating,li2016intelligence,li2018artificial,li2019parallel,riedmaier2020survey}, where safety-critical testing scenarios are usually purposely generated utilizing prior knowledge of CAVs such as surrogate models (SMs) of CAVs. However, due to the high complexity and black-box properties of CAVs, there exist significant performance dissimilarities between SMs and CAVs under test, which could severely compromise the effectiveness of the generated testing scenarios and decrease the evaluation efficiency.

Towards addressing this problem, several adaptive testing and evaluation methods have been proposed \cite{mullins2018adaptive,koren2018adaptive,feng2022adaptive,sun2021adaptive}. The basic idea of existing methods is to adaptively generate the testing scenarios during the testing process of CAVs. With more testing results of CAVs, more posteriori knowledge of CAVs can be obtained, and therefore the testing scenarios can be more customized and optimized for the CAVs under test. However, most existing methods can only be applied to relatively simple scenarios, and how to handle high-dimensional scenarios remains an open question. For example, Mullins \textit{et al}. \cite{mullins2018adaptive} proposed an adaptive sampling method that uses Gaussian process regression (GPR) and $k$-nearest neighbors to discover performance boundaries of the system under test and then updates the SM with new testing results obtained near the performance boundaries. Koren \textit{et al}. \cite{koren2018adaptive} put forward an adaptive stress testing method that uses deep reinforcement learning to find the most-likely failure scenarios. Feng \textit{et al}. \cite{feng2022adaptive} proposed an adaptive testing scenario library generation method using Bayesian optimization techniques with classification-based GPR and acquisition functions to select subsequent testing scenarios and then update the SMs with new testing results. Sun \textit{et al}. \cite{sun2021adaptive} presented an adaptive design of experiments method to detect safety-critical scenarios, which uses supervised machine learning models as SMs to approximate the testing results and devises acquisition functions for updating the SMs. 

The challenge for adaptively generating high-dimensional scenarios comes from the compounding effects of the ``Curse of Rarity'' (CoR) and the ``Curse of Dimensionality'' (CoD) \cite{liu2022curse}. The CoR refers to the concept that, due to rarity of safety-critical events, the amount of data needed to obtain sufficient information grow dramatically, while the CoD refers to the dimensionality of variables to represent realistic scenarios, which makes the computation cost increase exponentially with the growth of scenario dimensions. Most existing scenario-based testing approaches can only handle short scenario segments with limited background road users, where the decision variables are low-dimensional, which cannot represent the full complexity and variability of the real-world driving environment \cite{feng2020safety,feng2021parti,feng2021partii,zhao2016accelerated,zhao2017accelerated}. Towards addressing this challenge, the naturalistic and adversarial driving environment (NADE) method has been developed in our previous work \cite{feng2021intelligent}, which can generate high-dimensional highway driving scenarios. However, the NADE did not consider the performance gap between CAVs and SMs, which could also slow down the testing process. To the best of the authors' knowledge, there is no existing work that can handle the adaptive testing and evaluation problem in high-dimensional scenarios, and the goal of this paper is to fill this gap.

\begin{figure}[!t]
  \centering
  \includegraphics[width=8.85cm]{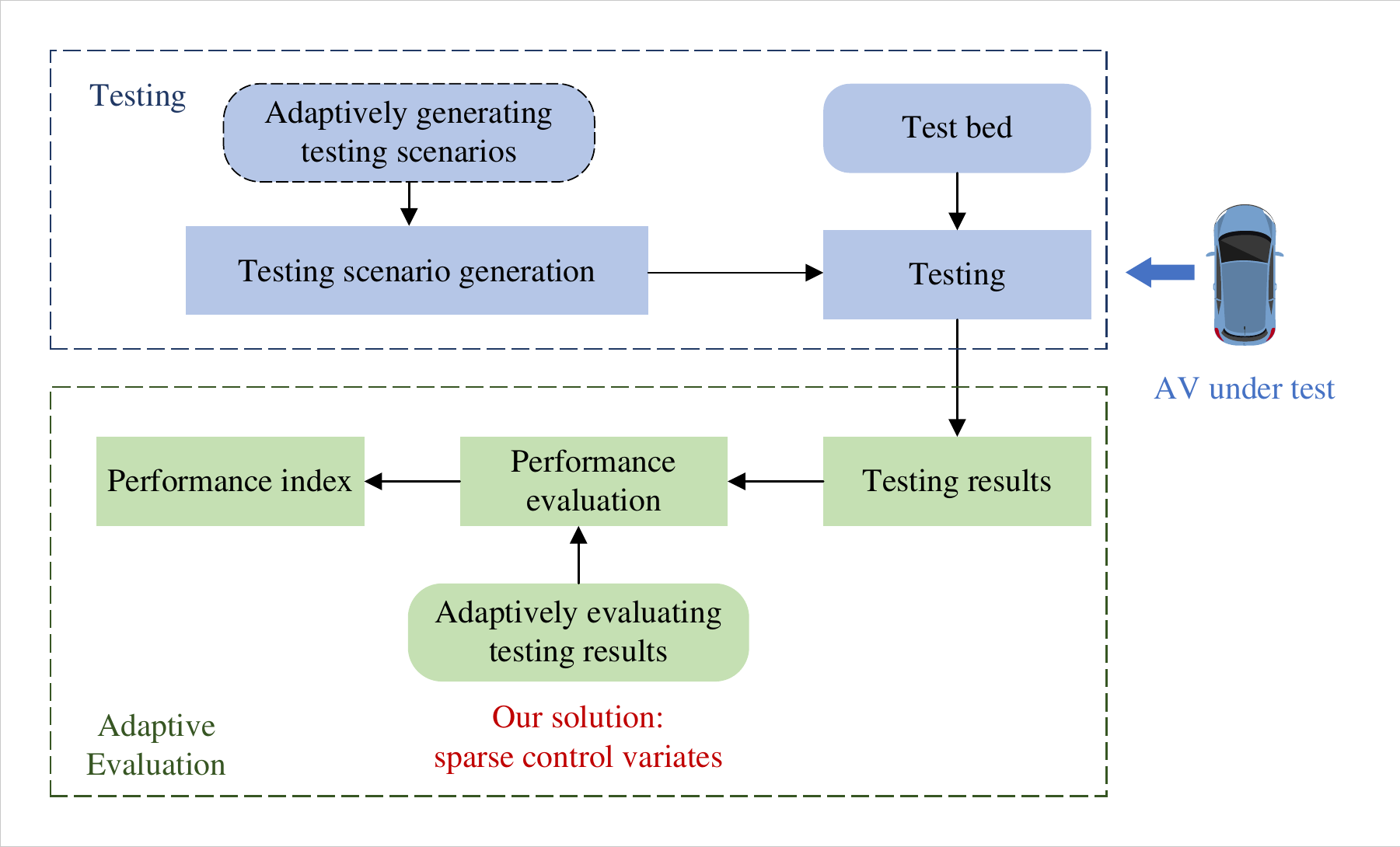}
  \caption{Illustration of the adaptive testing and evaluation framework. The focus of this study is the adaptive evaluation method for high-dimensional scenarios, where the sparse control variates method is proposed.}
  \label{fig:atscv_framework}
\end{figure}

In general, the adaptive testing and evaluation methods can be categorized into two types including adaptive testing scenario generation and adaptive testing result evaluation, which are complementary to each other as shown in Fig.~\ref{fig:atscv_framework}. Most existing studies focus on the former one, while in this study, we focus on the latter one and propose an adaptive evaluation framework that can handle high-dimensional scenarios. We note that how to realize the former one in high-dimensional scenarios also remains unsolved, which we leave for future study. In the proposed framework, we apply the NADE method to generate high-dimensional testing scenarios, where combinations of multiple SMs are utilized to improve the robustness of the generated scenarios for different CAVs under test. Then we propose a sparse control variate (SCV) method to adjust the testing results and evaluate CAVs' performance adaptively. Essentially, the SCV method could reduce the estimation variance for the CAV under test and thus reduce the required number of tests, accelerating the evaluation process adaptively. 

In the following paragraphs, we further explain the major idea of the proposed SCV method. The control variates (CV) method \cite{rubinstein1985efficiency} is a popular variance reduction technique applied in research areas such as deep learning \cite{grathwohl2018backpropagation} and reinforcement learning \cite{cheng2020trajectory}. Suppose we want to estimate $\mu\triangleq\mathbb{E}_p[f(X)]$ by Monte Carlo sampling \cite{shapiro2003monte}, where $p$ is the probabilistic distribution of the random variable $X$ and $f$ is the performance index of interest. Instead of directly estimating the unknown quantity $\mu$, the control variates method estimates the differences between the unknown quantity and known quantity as $\mu'\triangleq\mathbb{E}_p[f(X)-h(X)+\theta]$, where $h(X)$ is the control variate and $\theta\triangleq\mathbb{E}_p[h(X)]$ is a known value. Then, if $h(X)$ correlates with the performance index $f(X)$ (hence can provide some information about $f(X)$), the estimation variance of $\mu'$ will always be less than directly estimating $\mu$ \cite{owen2013monte}.
For testing and evaluation of CAVs, the control variate $h(X)$ can be designed by utilizing the prior knowledge of CAVs (e.g., different SMs). $h(X)$ usually contains adjustable control parameters, which can be optimized by leveraging the testing results. In such way, the information about the CAV under test could be incorporated, which makes the adaptive evaluation possible. However, due to the CoD, the computation cost of optimal control parameters will increase exponentially with the growth of scenario dimensions, so directly applying the ordinary CV method in high-dimensional scenarios is problematic.

\begin{figure}[!t]
  \centering
  \includegraphics[width=8.85cm]{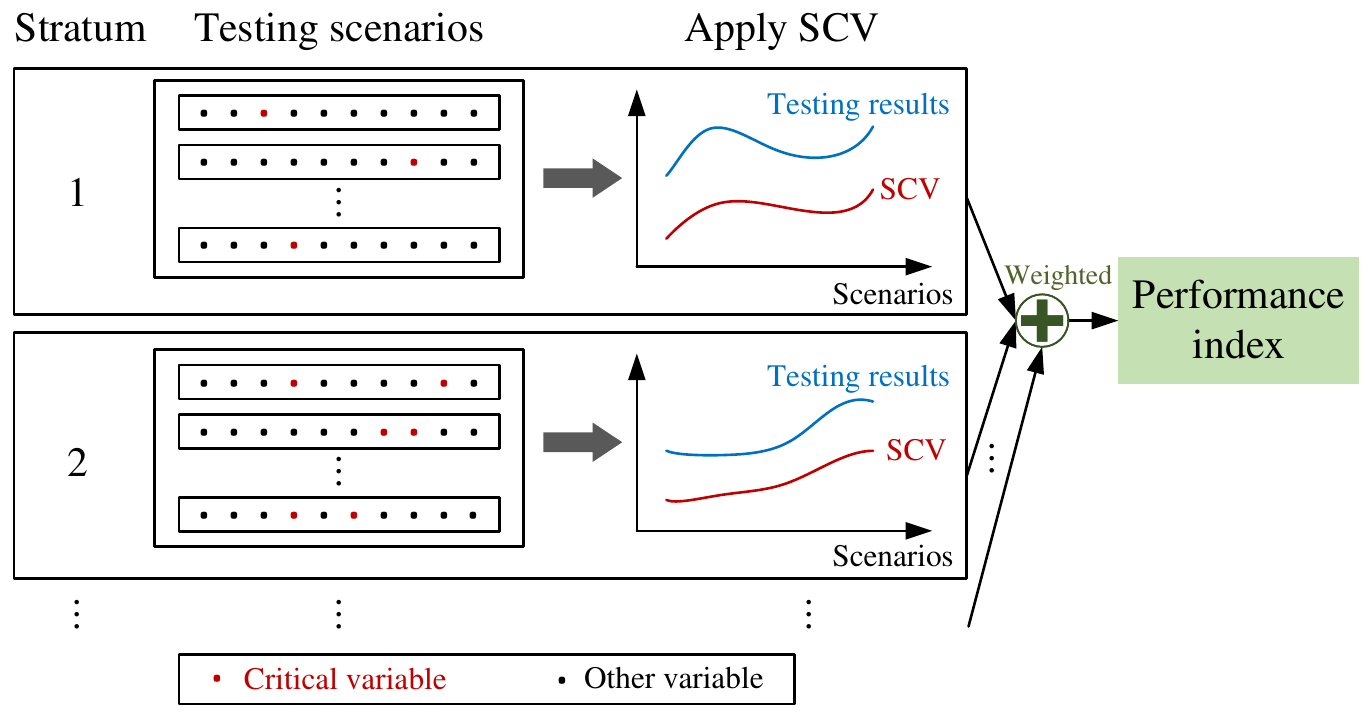}
  \caption{Illustration of the sparse control variates method. The SCV are constructed by only considering critical variables (represented as red dots in testing scenarios). The testing results are stratified into strata according to the number of critical variables and then adjusted by SCV within each stratum. Finally, the performance index are obtained by summing up these evaluation results with proportion weights.}
  \label{fig:scv_method}
\end{figure}

To address this problem, we propose the sparse control variates (SCV) method, as shown in Fig.~\ref{fig:scv_method}. The key idea is to construct the SCV by only considering the sparse but critical variables (e.g., behaviors of principal other vehicles at critical moments), following the similar idea from \cite{feng2021intelligent} that handles the CoD. However, the number of critical variables varies in different testing scenarios, which cannot be handled by ordinary CV method. To address this issue, in the SCV method, we stratify the testing scenarios into strata according to the number of critical variables. Then the control parameters can be optimized by multiple linear regression (MLR) \cite{olive2017multiple} within each stratum, and the final evaluation results are obtained by summing up those evaluation results in each stratum with the proportion weights. Since the number of critical variables is much less than the dimension of testing scenarios, the computation cost of optimal control parameters for SCV could be greatly reduced, overcoming the CoD challenge.

To verify the proposed method, we theoretically analyze its accuracy, efficiency, and optimality. The theorems show that our method is unbiased, and its estimation variance is nearly proportional to the best one that all the SMs used for generating testing scenarios could have. Moreover, under certain assumptions about the SMs, our method can provide a zero-variance estimator. To validate our method, the high-dimensional overtaking scenarios with large-scale naturalistic driving data are investigated. Simulation results show that our method can further accelerate the evaluation process by about one order of magnitude for different types of CAV models, comparing with the estimation efficiency in NADE.

Compared with our previously published conference paper about SCV \cite{yang2022adaptive}, the new contributions of this paper are listed as follows. First, we significantly extend our methodology into high-dimensional scenarios and establish the theoretical analysis for the accuracy, efficiency, and optimality of the proposed method with rigorous proofs. Second, a more realistic overtaking case study with large-scale naturalistic driving data is investigated to systematically validate the performances of our method. 

The remainder of this paper is organized as follows. Section \ref{sec:preliminaries} provides preliminary knowledge for the generation of NDE and NADE. Section \ref{sec:problem_formulation} formulates the adaptive testing and evaluation problem and elaborates the challenges of applying ordinary CV for adaptive safety evaluation. To address these challenges, in Section \ref{sec:ATSCV}, the SCV method is proposed. Then Section \ref{sec:theoretical_analysis} and \ref{sec:case_study} verify and validate the accuracy and efficiency of the proposed method from the theoretical and experimental perspectives, respectively. Finally, Section \ref{sec:conclusion} concludes the paper and discusses future research.

\section{Preliminaries}
\label{sec:preliminaries}

\subsection{Naturalistic Driving Environment Testing}

As discussed above, the prevailing approach for CAV evaluation is to test CAVs in the naturalistic driving environments (NDE) \cite{yan2021distributionally}, observe their performances, and make statistical comparisons with human drivers. In NDE, one of the vehicles is the automated vehicle (AV) under test and the others are background vehicles (BVs), which can be formulated as Markov games \cite{lowe2017multi}. A Markov game for $N$ agents (i.e., BVs) is defined by a set of states $\mathcal{S}$ describing the positions and velocities of all vehicles and a collection of action (i.e., acceleration) sets $\mathcal{A}_1,\dots,\mathcal{A}_N$, one for each agent in NDE. The total action space is denoted as $\mathcal{A}=\mathcal{A}_1\times\cdots\times\mathcal{A}_N$. Then a scenario is defined as the time series of the states of all vehicles and the actions of all agents, i.e.,
\begin{equation}
  x=(s_0,a_0,\dots,s_T,a_T)\in\mathcal{X},
\end{equation}
where $x$ represents the scenario, $\mathcal{X}$ is the set of all feasible scenarios, $s_t\in\mathcal{S}$ is the state of all vehicles at time $t$, $a_t\in\mathcal{A}$ is the action of all agents at time $t$, and $T$ is the time horizon.

Let $\Omega =\mathcal{X}$ be the sample space incorporating all feasible scenarios. Consider the probability space $(\Omega, \mathcal{F}, \mathbb{P})$, where $\mathcal{F}\triangleq2^\Omega$ is the power set of $\Omega$ and $\mathbb{P}$ is a probability measure on $\mathcal{F}$. Let $X:x\mapsto x$, $\forall x\in\mathcal{X}$ be the random variable of scenarios. For testing and evaluation of CAVs, the crash event is usually of most interest, which can be defined as $A=\{x\in\mathcal{X}:s_T\in\mathcal{S}_c\}$, where $\mathcal{S}_c$ is the set of all crash states. Then the crash rate is selected as the performance index, which can be computed as
\begin{equation}
  \mu=\mathbb{P}(A)
  =\mathbb{E}_p[\mathbb{I}_A(X)]
  =\sum_{x\in\mathcal{X}}\mathbb{P}(A|x)p(x),
\end{equation}
where $\mathbb{I}_A$ is the indicator function of $A$, and $p$ is the naturalistic joint distribution of $x$. The essence of testing AV in NDE is to estimate the performance index $\mu$ by Monte Carlo simulation, i.e.,
\begin{equation}
  \hat{\mu}_n = \frac{1}{n}\sum_{i=1}^n\mathbb{P}(A|X_i), \quad X_i\sim p.
\end{equation}

\subsection{Naturalistic and Adversarial Driving Environment Generation}

The NDE faces the CoR, making its estimation catastrophically inefficient. To improve the estimation efficiency, the importance sampling (IS) technique \cite{zhao2016accelerated,feng2021parti,feng2021partii} has been used to sample testing scenarios from the importance function $q$, which puts more weights on crash-prone scenarios. In IS, the performance index can be estimated as
\begin{equation}
  \hat{\mu}_q
  =\frac{1}{n}\sum_{i=1}^n\frac{\mathbb{P}(A|X_i)p(X_i)}{q(X_i)}, \quad X_i\sim q.
\end{equation}
However, the IS method faces the CoD if the testing scenarios are high-dimensional \cite{au2003important}. To address both the CoR and the CoD, the naturalistic and adversarial driving environment (NADE) \cite{feng2021intelligent} has been proposed to only sample critical variables of testing scenarios from importance functions, while other variables remain their naturalistic distributions.

Denote $x=(x_c, x_{-c})$, where $x_c=\{x_{c_1},\dots,x_{c_l}\}$ is the set of critical variables, $c_1,\dots,c_l$ are called the critical moments, $l=0,1,\dots,L$ is the number of control steps (i.e., the number of critical variables in $x_c$), and $x_{-c}$ is the set of other variables. Let $X_c:x\mapsto x_c$ be the random variable of critical variables and $X_{-c}:x\mapsto x_{-c}$ be the random variable of other variables, then we have $X=(X_c, X_{-c})$. The importance function can then be formulated as $q(x)=q(x_c)p(x_{-c})$, and therefore the performance index can be estimated in NADE as
\begin{equation}
  \tilde{\mu}_q
  =\frac{1}{n}\sum_{i=1}^n\frac{\mathbb{P}(A|X_i)p(X_{c,i})}{q(X_{c,i})}, \quad X_i\sim q,
\end{equation}
where $X_{c,i}$ is the random variable of critical variables of $X_i$.

\section{Problem Formulation}
\label{sec:problem_formulation}

\subsection{Adaptive Testing and Evaluation}

Due to the black-box property and various types of CAVs, how to adaptively test and evaluate CAVs remains a major challenge. One way of adaptive testing and evaluation is adaptively generating testing scenarios. For example, we can minimize the estimation variance by optimizing the importance function, i.e.,
\begin{equation}
  \min_{q\in\mathcal{Q}}~\mathrm{Var}_q\left(\frac{\mathbb{P}(A|X)p(X)}{q(X)}\right),
\end{equation}
where $\mathcal{Q}$ is the function space of $q$. Better importance functions can be found by leveraging the posteriori knowledge of CAVs obtained from testing results. Then the testing scenarios can be adaptively generated by sampling from updated importance functions.

In this paper, we focus on another way of adaptive testing and evaluation, i.e., adaptively evaluating weighted testing results. Specifically, the control variates (CV) method is adopted. This problem can be formulated as
\begin{equation}
  \min_{h\in\mathcal{H}}~\mathrm{Var}_q\left(\frac{\mathbb{P}(A|X)p(X)}{q(X)}-h(X)\right),
\end{equation}
where $h:\mathcal{X}\to\mathbb{R}$ is the control variate and $\mathcal{H}$ is the function space of $h$. The goal is to further reduce the estimation variance by optimizing $h$ in $\mathcal{H}$, leveraging the testing results.

\subsection{Control Variates}
\label{subsec:control_variates}

Control variates are widely used as a basic variance reduction technique in Monte Carlo simulation. They can be usefully combined with the mixture importance sampling, where individual importance functions can serve as CV. In mixture IS, the scenarios $X_i, i=1,\dots,n$ are sampled from the mixture importance function $q_\alpha=\sum_{j=1}^J\alpha_jq_j$, where $\alpha_j\geqslant0$, $\sum_{j=1}^J\alpha_j=1$ and the $q_j$ are importance functions. One commonly used way is to construct CV by using the linear combination of individual importance functions as
\begin{equation}
  \label{eq:ordinary_CV}
  h_\beta(X)=\sum_{j=1}^J\beta_j\left[\frac{q_j(X)}{q_\alpha(X)}-1\right],
\end{equation}
where $\beta=(\beta_1,\dots,\beta_J)^\top$ is the control vector, $\beta_j\in\mathbb{R}$ are control parameters, and $q_j/q_\alpha-1$ are individual control variate. Combining the control variate $h_\beta$ with mixture IS gives the estimation
\begin{equation}
  \label{eq:MIS_CV}
  \hat{\mu}_{q_\alpha,\beta}=\frac{1}{n}\sum_{i=1}^n\left[\frac{\mathbb{P}(A|X_i)p(X_i)}{q_\alpha(X_i)}-h_\beta(X_i)\right]
\end{equation}
for $X_i\sim q_\alpha$.

The unbiasedness of $\hat{\mu}_{q_\alpha,\beta}$ is guaranteed since
\begin{equation}
  \mathbb{E}_{q_\alpha}[\hat{\mu}_{q_\alpha,\beta}]=\mathbb{E}_{q_\alpha}\left[\frac{\mathbb{P}(A|X)p(X)}{q_\alpha(X)}-h_\beta(X)\right]=\mu,
\end{equation}
where the second equality is obtained from the unbiasedness of IS and $\mathbb{E}_{q_\alpha}[h_\beta(X)]=0$. The variance of $\hat{\mu}_{q_\alpha,\beta}$ can be compared to that of IS with individual importance functions $q_j$. We have the following lemma.

\begin{lemma}
  \label{lem:MIS_CV}
  Let $\beta^*$ be any minimizer over $\beta$ of $\mathrm{Var}_{q_\alpha}(\hat{\mu}_{q_\alpha,\beta})$, then
  \begin{equation}
    \mathrm{Var}_{q_\alpha}(\hat{\mu}_{q_\alpha,\beta^*})\leqslant
    \min_{1\leqslant j\leqslant J} \frac{\sigma_{q_j}^2}{n\alpha_j},
  \end{equation}
  where $\sigma_{q_j}^2$ is the asymptotic variance of $\hat{\mu}_{q_j}$, i.e.,
  \begin{equation}
    \sigma_{q_j}^2=
    \mathrm{Var}_{q_j}\left(\frac{\mathbb{P}(A|X)p(X)}{q_j(X)}\right),~j=1,\dots,J.
  \end{equation}
\end{lemma}

\begin{IEEEproof}
  This is the Theorem 2 in \cite{owen2000safe}.
\end{IEEEproof}

It can be seen from Lemma \ref{lem:MIS_CV} that the variance of $\hat{\mu}_{q_\alpha,\beta}$ will be zero if any one of the $q_j$ is optimal. This is a significant feature because we can nearly omit the influence of all other worse-performed importance functions. In applications, using only one SM to test CAVs is usually under huge risk, because the performance gap between the SM and various types of CAVs may be too large to give a good estimation efficiency. Therefore, to ensure the robustness, we can combine multiple SMs to test the CAVs. However, there often exist some poor-performed SMs that will compromise the overall estimation efficiency. Using mixture IS with CV provides an effective way to ensure both good estimation efficiency and robustness to various types of CAVs. 

In practice, the optimal control vector $\beta^*$ is usually unknown, and its estimation $\hat{\beta}$ can be obtained by multiple linear regression (MLR). Denote the weighted testing results as $Y_i=\mathbb{P}(A|X_i)p(X_i)/q_\alpha(X_i)$, $i=1,\dots,n$, and the individual control variate as $Z_{ij}=q_j(X_i)/q_\alpha(X_i)-1$, $i=1,\dots,n$, $j=1,\dots,J-1$. Then the $\hat{\beta}$ is given as the vector of coefficients obtained from MLR of $Y_i$ on $Z_{ij}$. In essence, this process is to search for the best control variate defined in Eq.~(\ref{eq:ordinary_CV}) in the function space spanned by individual control variate $q_j/q_\alpha-1$. However, challenges of estimating optimal control parameters arise when the testing scenarios are high-dimensional.

\subsection{CoD of Control Variates}
\label{subsec:CoD_of_CV}

Considering the Markov chain structure of scenarios with $T+1$ time steps, the mixture importance function is given by
\begin{equation}
  q_\alpha(x)=q_\alpha(s_0)\prod_{t=0}^T q_\alpha(a_t|s_t),~\forall x\in\mathcal{X},
\end{equation}
where $q_{\alpha}(s)=\sum_{j=1}^J {\alpha_jq_j(s)},~\forall s\in\mathcal{S}$, and $q_{\alpha}(a|s)
=\sum_{j=1}^J{\alpha_jq_j(a|s)},~\forall a\in\mathcal{A},~s\in\mathcal{S}$. It can be found that $q_\alpha(x)$ is the product of $T+2$ individual importance functions and thus is also the summation of $J^{T+2}$ combinations of different importance functions at each time step. Specifically, these individual importance functions are
\begin{equation}
  q_{j_0,\dots,j_{T+1}}(x)=
  q_{j_0}(s_0)q_{j_1}(a_0|s_0)\cdots q_{j_{T+1}}(a_T|s_T),
\end{equation}
where $j_0,\dots,j_{T+1}=1,\dots,J$. Then the individual control variate are given by $q_{j_0,\dots,j_{T+1}}/q_\alpha-1$. 

To find the estimation of optimal control parameters, we have to conduct MLR of $n$ weighted testing results on $J^{T+2}$ individual control variate. The number $J^{T+2}$ will increase exponentially with the dimension of scenarios, leading to the CoD of MLR. For example, if we have $J=10$ individual importance functions and the testing scenarios last for 10 seconds at a frequency of 10 Hz, then the number of individual control variate will be 10\textsuperscript{102}. This means that a matrix with dimension 10\textsuperscript{102} should be inverted in MLR, which is not tractable. Moreover, the situation will get even worse if the duration of scenarios grows to several hours, which are common in daily driving yet far from being tractable. The following section aims to address this challenge.

\section{Adaptive Safety Evaluation with Sparse Control Variates}
\label{sec:ATSCV}

In this section, we will address the CoD discussed above and show how to estimate the optimal control parameters. 

\subsection{Sparse Control Variates}
\label{subsec:scv}

We propose the sparse control variates (SCV) method to address the CoD of applying CV in high-dimensional scenarios. Specifically, the SCV are constructed by only considering the importance functions of only sparse and critical variables in high-dimensional testing scenarios. The number of critical variables is usually much less than the dimension of scenarios in NADE. Therefore, the number of SCV is also much less than the number of ordinary CV, which could greatly address the CoD. However, as the number of SCV varies in different testing scenarios, we can not directly apply SCV to the weighted testing results. Towards addressing this issue, we propose to stratify the testing scenarios into strata according to the number of critical variables and then apply SCV within each stratum.

Let $\mathcal{X}_l=\{x\in\mathcal{X}:|x_c|=l\}$, $l=0,1,\dots,L$ be the stratum of scenarios that are controlled $l$ steps, satisfying $\bigcup_{l=0}^L\mathcal{X}_l=\mathcal{X}$. Using mixture importance function $q_\alpha$, the estimation of the performance index in NADE is
\begin{equation}
  \label{eq:MIS_NADE}
  \tilde{\mu}_{q_\alpha}
  =\frac{1}{n}\sum_{i=1}^n\frac{\mathbb{P}(A|X_i)p(X_{c,i})}{q_\alpha(X_{c,i})}, \quad X_i\sim q_\alpha.
\end{equation}
The performance index of scenarios in stratum $\mathcal{X}_l$ can be written as $\mu_l\triangleq\mathbb{E}_p[\mathbb{I}_A(X)\mathbb{I}_{\mathcal{X}_l}(X)]$, $l=0,1,\dots,L$, then we have
\begin{equation}
  \mu=\sum_{l=0}^L\mathbb{E}_p[\mathbb{I}_A(X)\mathbb{I}_{\mathcal{X}_l}(X)]=\sum_{l=0}^L\mu_l.
\end{equation}
Similar to Eq.~(\ref{eq:MIS_NADE}), the estimation of $\mu_l$ is given by
\begin{equation}
  \label{eq:mu_est_stratum}
  \tilde{\mu}_{l,q_\alpha}
  =\frac{1}{n}\sum_{i=1}^n\frac{\mathbb{P}(A|X_i)\mathbb{I}_{\mathcal{X}_l}(X_i)p(X_{c,i})}{q_\alpha(X_{c,i})},
\end{equation}
and then we have
\begin{equation}
  \begin{aligned}
    \tilde{\mu}_{q_\alpha}
    &=\sum_{l=0}^L\frac{1}{n}\sum_{i=1}^n\frac{\mathbb{P}(A|X_i)\mathbb{I}_{\mathcal{X}_l}(X_i)p(X_{c,i})}{q_\alpha(X_{c,i})}
    =\sum_{l=0}^L\tilde{\mu}_{l,q_\alpha}.
  \end{aligned}
\end{equation}

Let $q_{j_1,\dots,j_l}(x)=p(x_{-c})q_{j_1}(x_{c_1})\cdots q_{j_l}(x_{c_l})$ be the importance functions that sample $x_{-c}$ from $p$ and sample $x_{c_1},\dots,x_{c_l}$ from $q_{j_1},\dots,q_{j_l}$ respectively, where $j_1,\dots,j_l=1,\dots,J$, $l=1,\dots,L$. Then the individual importance functions of critical variables are given by $q_{j_1,\dots,j_l}(x_c)$. Denote the linear combination of these individual importance functions as
\begin{equation}
  \tilde{h}_l(x)\triangleq
  \sum_{j_1,\dots,j_l}\beta_{l,j_1,\dots,j_l}q_{j_1,\dots,j_l}(x),~l=1,\dots,L,
\end{equation}
where $\beta_{l,j_1,\dots,j_l}\in\mathbb{R}$ are associated control parameters. Then the SCV are given by
\begin{equation}
  h_l(x_c)=\frac{\tilde{h}_l(x_c)\mathbb{I}_{\mathcal{X}_l}(x_c)}{q_\alpha(x_c)}-\theta_l,~l=1,\dots,L,
\end{equation}
where $\theta_l\triangleq\mathbb{E}_{q_\alpha}\big[\tilde{h}_l(X)\mathbb{I}_{\mathcal{X}_l}(X)/q_\alpha(X)\big]$. 
Therefore, the estimation $\tilde{\mu}_{l,q_\alpha}$ in Eq.~(\ref{eq:mu_est_stratum}) can be evaluated with SCV as
\begin{equation}
  \label{eq:ATSCV}
  \begin{aligned}
    \tilde{\mu}_{l,q_\alpha,\beta_l}
    &=\frac{1}{n}\sum_{i=1}^n\left[\frac{\mathbb{P}(A|X_i)\mathbb{I}_{\mathcal{X}_l}(X_i)p(X_{c,i})}{q_\alpha(X_{c,i})}-h_l(X_{c,i})\right]\\
    &=\frac{1}{n}\sum_{i=1}^n\frac{\mathbb{P}(A|X_i)p(X_{c,i})-\tilde{h}_l(X_{c,i})}{q_\alpha(X_{c,i})}\mathbb{I}_{\mathcal{X}_l}(X_i)+\theta_l\\
  \end{aligned}
\end{equation}
for $l=1,\dots,L$, where $\beta_l=\mathrm{vec}(\beta_{l,j_1,\dots,j_l})$ is the vector of control parameters, and $\mathrm{vec}(\cdot)$ is the vectorization operator that flattens a tensor into a long vector. Note that there is no critical variable for $l=0$, and thus we set $\beta_0\triangleq0$. In summary, the performance index estimated by the proposed SCV method is given by
\begin{equation}
  \label{eq:estimation_scv}
  \tilde{\mu}_{q_\alpha,\beta}=
  \sum_{l=0}^L\tilde{\mu}_{l,q_\alpha,\beta_l},
\end{equation}
where $\beta=\{\beta_l\}_{l=0}^L$ is the set of all control vectors.

\subsection{Optimal Control Parameters}
\label{subsec:optimal_params}

To estimate the optimal control parameters that minimize the estimation variance, multiple linear regression (MLR) technique is applied in each stratum. Let $\mathbb{X}_l\triangleq\{X_i|X_i\in\mathcal{X}_l, i=1,\dots,n\}$ be the set of sampled scenarios with $l$ controlled steps, $n_l\triangleq\sum_{i=1}^n\mathbb{I}_{\mathcal{X}_l}(X_i)$ be the number of tests with $l$ controlled steps and $d_l\triangleq J^l$ be the number of SCV, $l=1,\dots,L$. Denote the vector of testing results as
\begin{equation}
  Y_{l}\triangleq\left[\frac{\mathbb{P}(A|X_i)p(X_i)}{q_\alpha(X_i)}~\mathrm{for}~X_i\in\mathbb{X}_l\right]\in\mathbb{R}^{n_l},
\end{equation}
the individual SCV as
\begin{equation}
  h'_{j_1,\dots,j_l}(x_c)
  =\frac{q_{j_1,\dots,j_l}(x_c)}{q_\alpha(x_c)}-\sum_{x_c\in\mathcal{X}_l}q_{j_1,\dots,j_l}(x_c),
\end{equation}
for $l=1,\dots,L$. Then the matrix of individual SCV can be formulated as
\begin{equation}
  H_{l}\triangleq\left[\mathrm{vec}\left(h'_{j_1,\dots,j_l}(X_{c,i})\right)~\mathrm{for}~X_i\in\mathbb{X}_l\right]\in\mathbb{R}^{n_l\times d_l},
\end{equation}
for $l=1,\dots,L$. Then the regression formula is given by $Y_l\approx\eta_l+H_l\beta_l$. The MLR of $Y_l$ on $H_l$ is to find the optimal solution of the following optimization problem, i.e.,
\begin{equation}
  \min_{\eta_l,\beta_l}~f(\eta_l,\beta_l)=\|Y_l-\eta_l-H_l\beta_l\|_2^2.
\end{equation}

Letting the partial derivatives of $f$ with respect to $\eta_l$ and $\beta_l$ both equal zero, we have $\hat{\eta}_l=1^\top Y_l/n_l$ and $\hat{\beta}_l=(H_l^\top H_l)^{-1}H_l^\top Y_l$, assuming that the control matrix $M_l\triangleq H_l^\top H_l\in\mathbb{R}^{d_l\times d_l}$ is invertible. Then the estimated performance index is $\hat{\mu}_l=n_l\hat{\eta}_l/n$. In practice the control matrix may often not be invertible, then we use singular value decomposition (SVD) \cite{wall2003singular} to compute the regression coefficients $\hat{\beta}_l$, and the rank of the control matrix is
\begin{equation}
  \label{eq:min_num_test_dimen}
  \mathrm{rank}(M_l)=\mathrm{rank}(H_l)\leqslant\min\{n_l,d_l\}.
\end{equation}
If $n_l<d_l$, then the control matrix $M_l$ will be singular and has utmost $n_l$ nonzero singular values. As the number of tests $n_l$ in $\mathbb{X}_l$ will not grow exponentially with the number of control steps $l$, the rank of the control matrix will also not, albeit the dimension $d_l=J^l$ of the control matrix increases exponentially with $l$. In conclusion, solving the optimal control parameters for SCV is tractable and will not face the CoD challenge. We will further demonstrate this in Subsection \ref{subsec:results}.

\section{Theoretical Analysis}
\label{sec:theoretical_analysis}

This section theoretically justifies the accuracy, efficiency and optimality of the proposed SCV method.

\subsection{Accuracy Analysis}

We first prove that the estimation is unbiased.

\begin{theorem}
  Let $\tilde{\mu}_{q_\alpha,\beta}$ be given by Eq.~(\ref{eq:estimation_scv}) where $q_\alpha>0$ whenever $\mathbb{P}(A|x)p(x)>0$, then $\mathbb{E}_{q_\alpha}[\tilde{\mu}_{q_\alpha,\beta}]=\mu$.
\end{theorem}

\begin{IEEEproof}
  To establish unbiasedness, write
  \begin{equation}
    \begin{aligned}
      \mathbb{E}_{q_\alpha}[\tilde{\mu}_{q_\alpha,\beta}]
      &=\mathbb{E}_{q_\alpha}\left[\sum_{l=0}^L\tilde{\mu}_{l,q_\alpha,\beta_l}\right]\\
      &=\sum_{l=0}^L\mathbb{E}_{q_\alpha}\left[\tilde{\mu}_{l,q_\alpha}-\frac{\tilde{h}_l(X)}{q_\alpha(X)}\mathbb{I}_{\mathcal{X}_l}(X)+\theta_l\right]\\
      &=\sum_{l=0}^L(\mu_l-\theta_l+\theta_l)
      =\mu.\\
    \end{aligned}
  \end{equation}
\end{IEEEproof}

\begin{remark}
  This theorem indicates that the estimation is unbiased if the control parameters $\beta$ are independent of the sample data. It's worth noting that in practice the control parameters are usually estimated by the sample data, which would bring a bias. However, that bias is ordinarily negligible (please see Section 8.9 in \cite{owen2013monte} for more discussions).
\end{remark}

\subsection{Efficiency Analysis}

Next, we evaluate the efficiency of the SCV method. The variance of the estimation $\tilde{\mu}_{q_\alpha,\beta}$ is $\mathrm{Var}_{q_\alpha}(\tilde{\mu}_{q_\alpha,\beta})=\sigma_{q_\alpha,\beta}^2/n$, where $\sigma_{q_\alpha,\beta}^2$ is the asymptotic variance of $\tilde{\mu}_{q_\alpha,\beta}$, i.e.,
\begin{equation}
  \sigma_{q_\alpha,\beta}^2
  =\mathrm{Var}_{q_\alpha}\left(\sum_{l=0}^L\frac{\mathbb{P}(A|X)p(X)-\tilde{h}_l(X)}{q_\alpha(X)}\mathbb{I}_{\mathcal{X}_l}(X)\right)
\end{equation}
for $X\sim q_\alpha$. Denote
\begin{equation}
  Z_l\triangleq\frac{\mathbb{P}(A|X)p(X)-\tilde{h}_l(X)}{q_\alpha(X)}\mathbb{I}_{\mathcal{X}_l}(X),~l=0,\dots,L,
\end{equation}
then the asymptotic variance $\sigma_{q_\alpha,\beta}^2$ can be expressed as
\begin{equation}
    \sigma_{q_\alpha,\beta}^2
    =\mathrm{Var}_{q_\alpha}\left(\sum_{l=0}^LZ_l\right)
    =\mathbb{E}_{q_\alpha}\left[\left(\sum_{l=0}^L\Big[Z_l-\mathbb{E}_{q_\alpha}[Z_l]\Big]\right)^2\right].
\end{equation}
Let $L'=L+1$, then by convexity of quadratic function and Jensen's inequality, we have
\begin{equation}
  \begin{aligned}
    \sigma_{q_\alpha,\beta}^2
    &\leqslant\mathbb{E}_{q_\alpha}\left[L'\sum_{l=0}^L\Big(Z_l-\mathbb{E}_{q_\alpha}[Z_l]\Big)^2\right]\\
    &=L'\sum_{l=0}^L\mathrm{Var}_{q_\alpha}(Z_l).
  \end{aligned}
\end{equation}
Denote $\sigma_{l,q_\alpha,\beta_l}^2\triangleq\mathrm{Var}_{q_\alpha}(Z_l)$ and the asymptotic variance of $\tilde{\mu}_{l,q}$ over $\mathcal{X}_l$ as $\sigma_{l,q}^2$, i.e.,
\begin{equation}
  \sigma_{l,q}^2\triangleq\sum_{x\in\mathcal{X}_l}\left(\frac{\mathbb{P}(A|x)p(x)}{q(x)}-\mu_l\right)^2q(x),~l=1,\dots,L,
\end{equation}
then we have the following theorem.

\begin{theorem}
  If $\beta^*$ is any minimizer of $\sigma_{q_\alpha, \beta}^2$, then
  \begin{equation}
    \label{eq:thm_1}
    \begin{aligned}
      \sigma_{q_\alpha, \beta^*}^2
      &\leqslant L'\sigma_{0,p,\beta_0}^2 \\
      &\quad+ L'\sum_{l=1}^L\min_{j_1,\dots,j_l}\left\{\frac{\sigma_{l,q_{j_1,\dots,j_l}}^2}{\prod_{\ell=1}^l\alpha_{j_\ell}}+3\left(\frac{\mu_l}{\prod_{\ell=1}^l\alpha_{j_\ell}}\right)^2\right\}.\\
    \end{aligned}
  \end{equation}
\end{theorem}

\begin{IEEEproof}
  Take $\sigma_{1,q_\alpha,\beta_1}^2$ as an example. Following the proof in \cite{owen2000safe}, we consider the particular vector $\beta_1$ having $\beta_{1,1}=0$ and $\beta_{1,j}=-\mu_1\alpha_j/\alpha_1$ for $j>1$. Let $r_1(x)\triangleq[\mathbb{P}(A|x)p(x)-\mu_1q_1(x)]\mathbb{I}_{\mathcal{X}_1}(x)$, then we have $\sum_{x\in\mathcal{X}}r_1(x)=\mu_1(1-\xi_1)$, where $\xi_1\triangleq\sum_{x\in\mathcal{X}_1}q_1(x)$, $\xi_1\in[0,1]$. Substituting these values, we find that for this $\beta_1$,
  \begin{equation}
    \begin{aligned}
      Z_1
      &=\frac{\mathbb{P}(A|X)p(X)-\tilde{h}_1(X)}{q_\alpha(X)}\mathbb{I}_{\mathcal{X}_1}(X)\\
      &=\frac{\mathbb{P}(A|X)p(X)-\mu_1q_1+\mu_1q_1-\tilde{h}_1(X)}{q_\alpha(X)}\mathbb{I}_{\mathcal{X}_1}(X)\\
      &=\frac{r_1(X)}{q_\alpha(X)}+\frac{\mu_1}{\alpha_1}\mathbb{I}_{\mathcal{X}_1}(X),\\
    \end{aligned}
  \end{equation}
  and $\mathbb{E}_{q_\alpha}[Z_1]=\mu_1\alpha_{1,1}/\alpha_1$, where $\alpha_{1,1}\triangleq\alpha_1+\sum_{j=2}^J\alpha_j\linebreak\sum_{x\in\mathcal{X}_1}q_j(x)$, $\alpha_{1,1}\in[0,1]$. Therefore, we have
  \begin{equation}
    \begin{aligned}
      \sigma_{1,q_\alpha,\beta_1}^2
      &=\mathbb{E}_{q_\alpha}\left[\Big(Z_1-\mathbb{E}_{q_\alpha}[Z_1]\Big)^2\right]\\
      &=\sum_{x\in\mathcal{X}}\left[\frac{r_1(x)}{q_\alpha(x)}+\frac{\mu_1}{\alpha_1}\Big(\mathbb{I}_{\mathcal{X}_1}(x)-\alpha_{1,1}\Big)\right]^2q_\alpha(x)\\
      &\triangleq V_{1,1}+V_{1,2}+V_{1,3},\\
    \end{aligned}
  \end{equation}
  where
  \begin{equation}
    \begin{aligned}
      V_{1,1}
      &\triangleq\sum_{x\in\mathcal{X}}\frac{r_1^2(x)}{q_\alpha(x)}
      =\sum_{x\in\mathcal{X}}\frac{[\mathbb{P}(A|x)p(x)-\mu_1q_1(x)]^2}{q_\alpha(x)}\mathbb{I}_{\mathcal{X}_1}(x)\\
      &\leqslant\sum_{x\in\mathcal{X}_1}\frac{[\mathbb{P}(A|x)p(x)-\mu_1q_1(x)]^2}{\alpha_1q_1(x)}
      =\frac{\sigma_{1,q_1}^2}{\alpha_1},
    \end{aligned}
  \end{equation}
  \begin{equation}
    \begin{aligned}
      V_{1,2}
      &\triangleq\sum_{x\in\mathcal{X}}\frac{2\mu_1r_1(x)(\mathbb{I}_{\mathcal{X}_1}(x)-\alpha_{1,1})}{\alpha_1}\\
      &=\frac{2\mu_1^2(1-\xi_1)(1-\alpha_{1,1})}{\alpha_1}
      \leqslant 2\left(\frac{\mu_1}{\alpha_1}\right)^2,
    \end{aligned}
  \end{equation}  
  and
  \begin{equation}
    \begin{aligned}
      V_{1,3}
      &\triangleq\sum_{x\in\mathcal{X}}\left[\frac{\mu_1(\mathbb{I}_{\mathcal{X}_1}(x)-\alpha_{1,1})}{\alpha_1}\right]^2q_\alpha(x)\\
      &\leqslant \sum_{x\in\mathcal{X}}\left(\frac{\mu_1}{\alpha_1}\right)^2q_\alpha(x)
      =\left(\frac{\mu_1}{\alpha_1}\right)^2.
    \end{aligned}
  \end{equation}

  Therefore, we conclude that
  \begin{equation}
    \sigma_{1,q_\alpha,\beta_1^*}^2
    \leqslant \sigma_{1,q_\alpha,\beta_1}^2
    \leqslant \frac{\sigma_{1,q_1}^2}{\alpha_1} + 3\left(\frac{\mu_1}{\alpha_1}\right)^2.
  \end{equation}
  By making similar arguments for $j=2,\dots,J$, we have
  \begin{equation}
    \label{eq:var_bound_1}
    \sigma_{1,q_\alpha,\beta_1^*}^2\leqslant
    \min_{j}\left\{\frac{\sigma_{1,q_j}^2}{\alpha_j}+3\left(\frac{\mu_1}{\alpha_j}\right)^2\right\}.
  \end{equation}
  It's straightforward to extend the proof for $l=2,\dots,L$, then Eq.~(\ref{eq:thm_1}) is established.
\end{IEEEproof}

\begin{remark}
  For $l=1$, we expect to get approximately $n_1\alpha_j$ scenarios in $\mathcal{X}_1$ from the importance function $q_j$. The quantity $\sigma_{1,q_j}^2/\alpha_j$ in Eq.~(\ref{eq:var_bound_1}) is the variance we would obtain from $n_1\alpha_j$ such scenarios alone. It is hard to imagine that we could do better in general, because when $\sigma_{1,q_j}^2=\infty$ for all but one of the mixture components it is guaranteed that those bad components do not make the estimation worse than what we would have had from the one good importance function. Moreover, if there exists an optimal importance function in $q_j$, then the minimum value of $\sigma_{1,q_j}^2/\alpha_j$ will be zero, which will greatly reduce the estimation variance. It should be noted that the upper bound for variance in Eq.~(\ref{eq:var_bound_1}) contains a residual term $3(\mu_1/\alpha_j)^2$, which is the cost for stratifying the scenarios.
\end{remark}

\subsection{Optimality Analysis}

Under the following assumptions, the estimation variance of the SCV method can be zero.

\begin{assumption}
  The scenarios in $\mathcal{X}_0$ will not be sampled by $q_\alpha$, i.e., $q_\alpha(x)=0$, $\forall x\in\mathcal{X}_0$.
\end{assumption}

\begin{assumption}
  The control policy satisfies $|x_c|=1$, i.e., the number of critical variable of all sampled scenarios is 1.
\end{assumption}

\begin{assumption}
  There exists an optimal control policy such that $\mathbb{P}(A|x_c)=\mathbb{P}(A|x)$, which means that the critical variable $x_c$ can totally dominate the crash probability.
\end{assumption}

\begin{assumption}
  There exists an optimal importance function among $q_j$. Without loss of generality, let $q_1$ be the optimal importance function, i.e., $q_1(x_c)\triangleq\mathbb{P}(A|x_c)p(x_c)/\mu$.
\end{assumption}

\begin{theorem}
  Under Assumptions 1, 2, 3 and 4, if $\beta^*$ is any minimizer of $\sigma_{q_\alpha, \beta}^2$, then $\sigma_{q_\alpha, \beta^*}^2=0$.
\end{theorem}

\begin{IEEEproof}
  From Assumptions 1 and 2, we know that all sampled scenarios will only be controlled once, i.e., $\mathcal{X}=\mathcal{X}_1$ and $\mu=\mu_1$, then
  \begin{equation}
    Z_1=\frac{r_1(X)}{q_\alpha(X)}+\frac{\mu_1}{\alpha_1}\mathbb{I}_{\mathcal{X}_1}(X)=\frac{r_1(X)}{q_\alpha(X)}+\frac{\mu_1}{\alpha_1},
  \end{equation}
  and $\mathbb{E}_{q_\alpha}[Z_1]=\mu_1\alpha_{1,1}/\alpha_1=\mu_1/\alpha_1$. Therefore, the asymptotic variance $\sigma_{1,q_\alpha,\beta_1}^2$ is
  \begin{equation}
    \label{eq:asym_var_1}
    \begin{aligned}
      \sigma_{1,q_\alpha,\beta_1}^2
      &=\mathbb{E}_{q_\alpha}\left[\Big(Z_1-\mathbb{E}_{q_\alpha}[Z_1]\Big)^2\right]\\
      &=\sum_{x\in\mathcal{X}}\frac{r_1^2(x)}{q_\alpha(x)}
      \leqslant\frac{\sigma_{1,q_1}^2}{\alpha_1}.\\
    \end{aligned}
  \end{equation}
  By Assumptions 3 and 4, we have $\mathbb{P}(A|x_c)=\mathbb{P}(A|x)$ and $q_1(x_c)=\mathbb{P}(A|x_c)p(x_c)/\mu$, then
  \begin{equation}
    \begin{aligned}
      \sigma_{1,q_1}^2
      &=\sum_{x\in\mathcal{X}_1}\left(\frac{\mathbb{P}(A|x)p(x)}{q_1(x)}-\mu_1\right)^2q_1(x)\\
      &=\sum_{x\in\mathcal{X}}\left(\frac{\mathbb{P}(A|x_c)p(x_c)}{q_1(x_c)}-\mu\right)^2q_1(x)
      =0.\\
    \end{aligned}
  \end{equation}
  Therefore, we conclude that $\sigma_{q_\alpha, \beta^*}^2=\sigma_{1,q_\alpha,\beta_1}^2=0$.
\end{IEEEproof}

\begin{remark}
  Assumption 1 suggests that the scenarios in $\mathcal{X}_0$ should not be sampled. Since there are no crash in these scenarios, they can not make any contribution to the estimation. Assumption 2 requires that the number of critical variable is 1, because stratifying scenarios into different strata leads to some residual terms (e.g., $3(\mu_1/\alpha_j)^2$ in Eq.~(\ref{eq:var_bound_1})) in estimation variance that can not be eliminated. Assumption 3 indicates that the critical variables should dominate the crash probability, since otherwise we may lose some critical information about the scenarios and obtain the suboptimal testing results. Assumption 4 requires that one of the importance functions should be optimal, together with Assumption 3 further reducing the asymptotic variances to zero. Although in practice these assumptions may not be fully satisfied, they could provide useful guidance for us to implement the SCV method.
\end{remark}

\begin{remark}
  The theorems in this section hold regardless of the specifics of SMs, which may be constructed by traditional traffic models or by neural networks.
\end{remark}

\section{Overtaking Case Study}
\label{sec:case_study}

\subsection{Overtaking Scenarios}

\begin{figure}[!t]
  \centering
  \includegraphics[width=8.85cm]{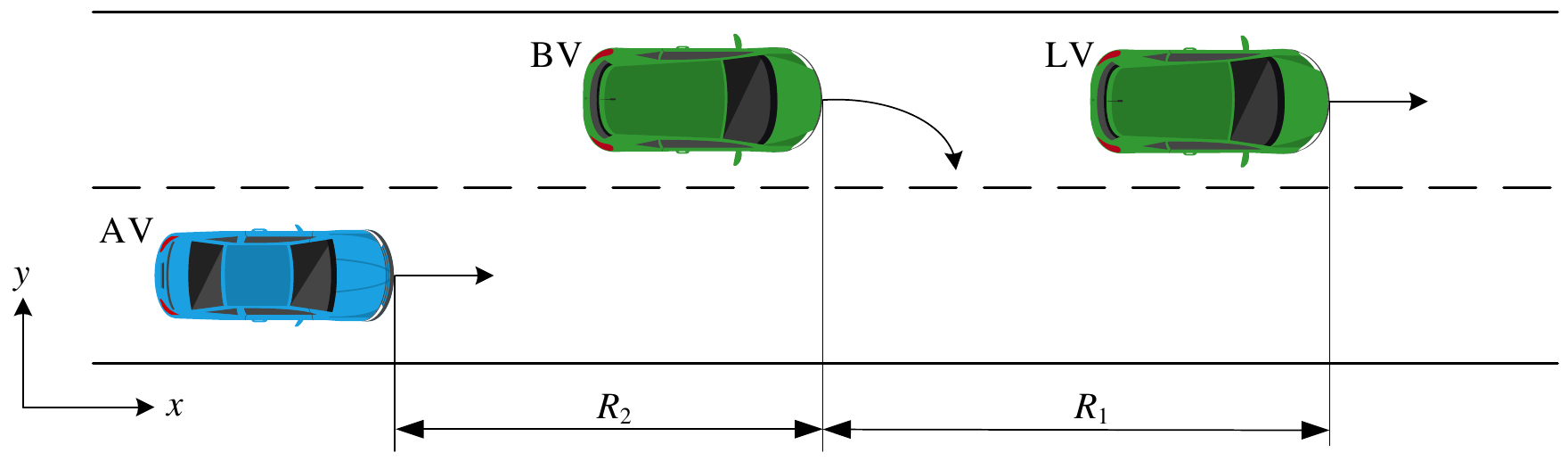}
  \caption{Illustration of the overtaking scenarios.}
  \label{fig:overtaking}
\end{figure}

The overtaking scenarios are shown in Fig.~\ref{fig:overtaking}, where the leading vehicle (LV) runs at the left lane, the background vehicle (BV) follows LV and the automated vehicle (AV) runs at the right lane. If BV cuts in to the right lane, then AV will follow BV and may rear-end BV, resulting in a crash. The state of the overtaking scenarios can be formulated as
\begin{equation}
  s\triangleq\big(v_{\mathrm{BV}}, R_1, \dot{R}_1, R_2, \dot{R}_2\big),
\end{equation}
where $R_1\triangleq x_{\mathrm{LV}}-x_{\mathrm{BV}}$, $\dot{R}_1\triangleq v_{\mathrm{LV}}-v_{\mathrm{BV}}$, $R_2\triangleq x_{\mathrm{BV}}-x_{\mathrm{AV}}$, and $\dot{R}_2\triangleq v_{\mathrm{BV}}-v_{\mathrm{AV}}$. The $x_{\mathrm{BV}}$, $x_{\mathrm{LV}}$, $x_{\mathrm{AV}}$ are the positions and $v_{\mathrm{BV}}$, $v_{\mathrm{LV}}$, $v_{\mathrm{AV}}$ are the velocities of BV, LV and AV, respectively. The action of the overtaking scenario is defined as the actions of LV and BV, i.e., $a\triangleq (a_{\mathrm{LV}},a_{\mathrm{BV}})$. We note that the overtaking scenarios are more stochastic and complicated than simple scenarios such as cut-in scenarios and car-following scenarios, since the BV in overtaking scenarios may have many chances to cut in, resulting in different cut-in scenarios and car-following scenarios between BV and AV. This is the reason why overtaking scenarios are always much more high-dimensional than cut-in scenarios. 

\subsection{Generation of NDE}

The essence of NDE is to provide a driving environment where all BVs travel like humans. To generate NDE, the probability distributions of the behaviors of all BVs should be consistent with the naturalistic driving data (NDD) \cite{yan2021distributionally}. In this paper, the probability distributions of free-driving, car-following, and cut-in behaviors are extracted from the NDD of the Safety Pilot Model Deployment (SPMD) \cite{bezzina2014safety} program and Integrated Vehicle-Based Safety System (IVBSS) \cite{sayer2011integrated} at the University of Michigan, Ann Arbor. The initial state is set as
\begin{equation}
  s_0=[v_{\mathrm{BV},0}, R_{1,0}, \dot{R}_{1,0}, R_{2,0}, \dot{R}_{2,0}],
\end{equation}
where $v_{\mathrm{BV},0}$, $R_{1,0}$, $\dot{R}_{1,0}$ are sampled from the naturalistic distributions of car-following scenarios, $R_{2,0}\sim\mathcal{U}(20~\text{m},100~\text{m})$, $\dot{R}_{2,0}\sim\mathcal{U}(-5~\text{m/s},-10~\text{m/s})$, where $\mathcal{U}$ is the uniform distribution. After sampling the initial state, all vehicles select actions independently and simultaneously for each time step (0.1 s). The cut-in maneuver of BV is set completed within one time step. The car-following maneuver of AV is controlled by the intelligent driver model (IDM)\cite{ro2017formal}. The simulation continues until AV rear-ends BV or maximum simulation time (20 s) reached. Typically, the dimension of overtaking scenarios will exceed 1400 (201 time steps, each with 5 state variables and 2 action variables), leading to the high-dimensionality challenge.

\subsection{Generation of NADE}

The goal of NADE is to generate high-dimensional testing scenarios where the behaviors of BVs are adjusted only at critical moments, while keeping naturalistic distributions as in NDE at other time steps \cite{feng2021intelligent}. To construct the importance function, the maneuver criticality of BV is evaluated at each time step, which is defined as the multiplication of the exposure frequency and the maneuver challenge. The exposure frequency represents the probability of each action given current state in NDE. The maneuver challenge measures the probability of crash between AV and BV given current state and action. Since the AV models are usually black-boxes, the surrogate models (SMs) are adopted to approximate the maneuver challenge. In this paper, we use IDM and full velocity difference model (FVDM) \cite{ro2017formal} as SMs with different parameters: (1) IDM, denoted as SM-I; (2) FVDM with $a_{\min}=-1$ m/s\textsuperscript{2}, denoted as SM-II; (3) FVDM with $a_{\min}=-6$ m/s\textsuperscript{2}, denoted as SM-III. Then the importance functions can be obtained from the maneuver criticalities estimated by these SMs. Readers can find more technical details in \cite{feng2021intelligent}.

\subsection{Application of SCV}

\begin{algorithm}[!t]
  \label{alg:ATSCV}
  \caption{Adaptive safety evaluation with sparse control variates by multiple linear regression}
  \KwIn{$p$, $q_\alpha$, $X_{c,i}$, and $\mathbb{P}(A|X_i)$, $i=1,\dots,n$}
  \KwOut{$\tilde{\mu}_{q_\alpha,\hat{\beta}}$, $\mathrm{Var}_{q_\alpha}(\tilde{\mu}_{q_\alpha,\hat{\beta}})$}
  initialize $Y_l$ and $H_l$ as empty arrays, $l=0,\dots,L$\;
  initialize $n_l=0$, $l=0,\dots,L$\;
  \For{$i\leftarrow 1$ \KwTo $n$}{
    $l\leftarrow$ number of control steps of $X_{c,i}$\;
    $n_l\leftarrow n_l+1$\;
    \eIf{$l=0$}{
      append $Y_l$ with $\mathbb{P}(A|X_i)$\;
      append $H_l$ with 0\;
    }{
      append $Y_l$ with $\mathbb{P}(A|X_i)p(X_{c,i})/q_\alpha(X_{c,i})$\;
      append $H_l$ with $\mathrm{vec}(q_{j_1,\dots,j_l}(X_{c,i})/q_\alpha(X_{c,i}))$, $j_1,\dots,j_l=1,\dots,J-1$\;
    }
  }
  \For{$l\leftarrow 0$ \KwTo $L$}{
    $H_l\leftarrow H_l-\mathrm{average}(H_l)$\;
    $\mathrm{MLR}\leftarrow$ multiple linear regression of $Y_l$ on $H_{l}$\;
    $\hat{\beta}_l\leftarrow$ estimated coefficients from $\mathrm{MLR}$\;
    $\hat{\eta}_l\leftarrow$ estimated intercept from $\mathrm{MLR}$\;
    $\tilde{\mu}_{l,q_\alpha,\hat{\beta}_l}\leftarrow n_l\hat{\eta}_l/n$, $Z_l\leftarrow Y_l-H_l\hat{\beta}_l$\;
  }
  $Z\leftarrow[Z_0,\dots,Z_L]$\;
  $\tilde{\mu}_{q_\alpha,\hat{\beta}}\leftarrow\sum_{l=0}^L\tilde{\mu}_{l,q_\alpha,\hat{\beta}_l}$, $\mathrm{Var}_{q_\alpha}(\tilde{\mu}_{q_\alpha,\hat{\beta}})\leftarrow\mathrm{var}(Z)$\;
  return $\tilde{\mu}_{q_\alpha,\hat{\beta}}$, $\mathrm{Var}_{q_\alpha}(\tilde{\mu}_{q_\alpha,\hat{\beta}})$\;
\end{algorithm}

As shown in Algorithm \ref{alg:ATSCV}, the SCV method can be applied to adjust the testing results and reduce estimation variance after testing AV in NADE. The key is to use importance functions of only sparse and critical variables to construct SCV, and then apply MLR of weighted testing results on SCV in each stratum. Finally, the estimated performance index is given by the summation of weighted intercepts obtained from MLR in all strata.

\subsection{Evaluation Results}
\label{subsec:results}

\begin{figure}[!t]
  \centering
  \includegraphics[width=7.7cm]{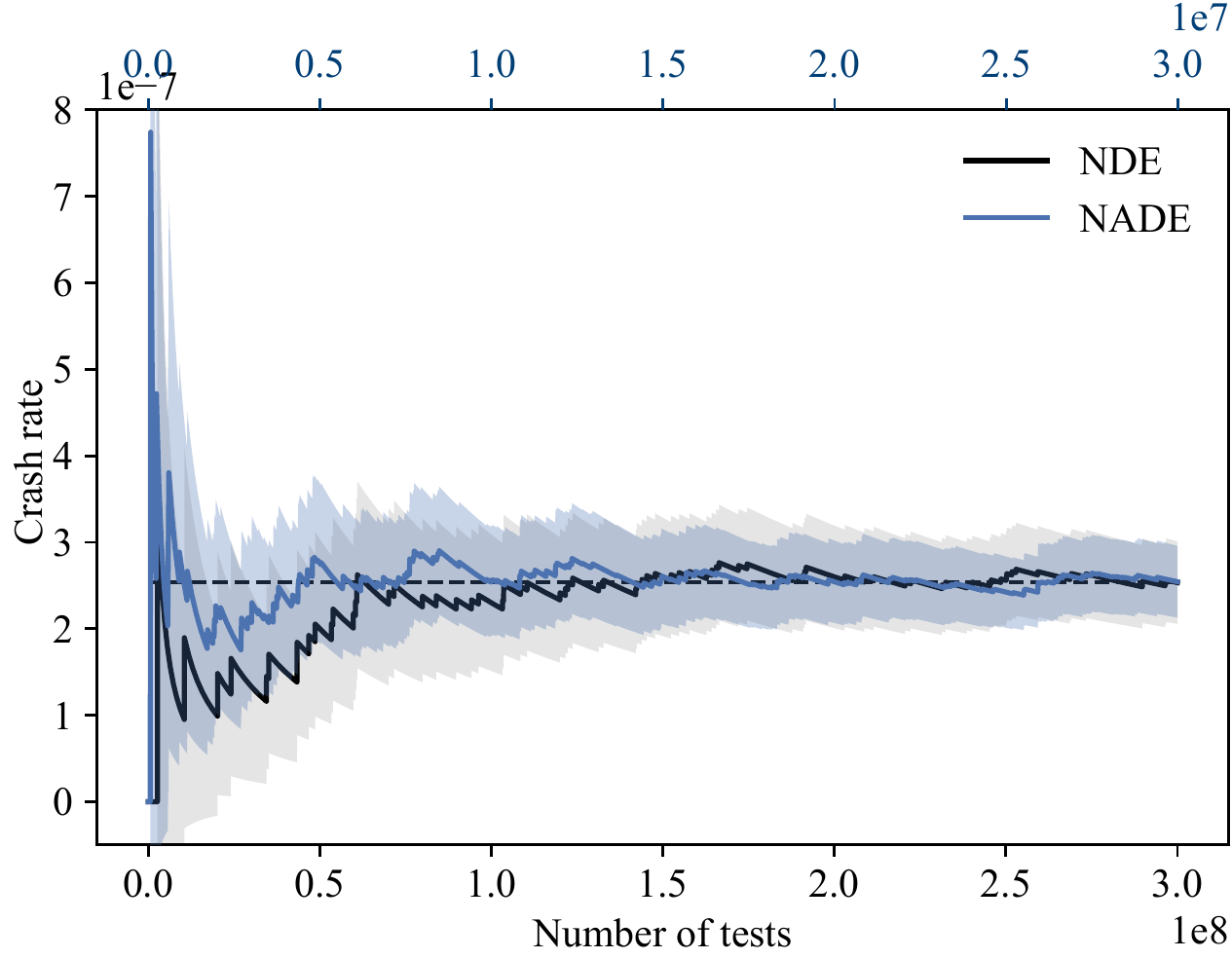}
  \caption{Crash rates of AV in NDE and NADE, where the dashed line is the crash rate estimated by NDE.}
  \label{fig:accident_rate_NDE_NADE}
\end{figure}

\begin{figure}[!t]
  \centering
  \includegraphics[width=8cm]{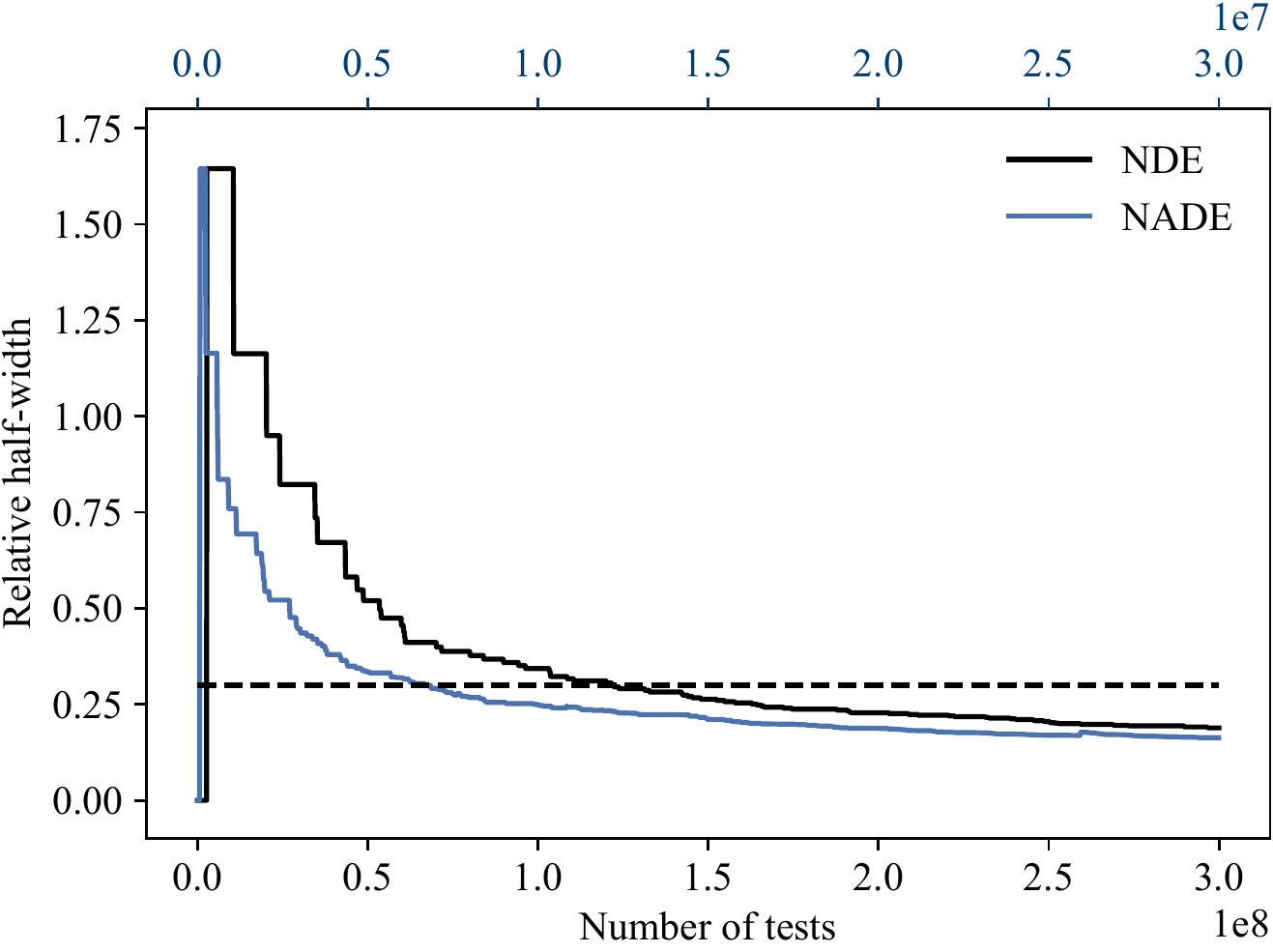}
  \caption{RHW of AV evaluation in NDE and NADE, where the dashed line represents the RHW threshold (0.3).}
  \label{fig:relative_half_width_NDE_NADE}
\end{figure}

We validate the accuracy and efficiency of AV evaluation in NDE and NADE by the simulation of overtaking scenarios. The simulation is parallel conducted using 100 threads on a computer equipped with AMD\textsuperscript{\textregistered} EPYC\texttrademark{} 7742 CPU and 512 GB RAM. Fig.~\ref{fig:accident_rate_NDE_NADE} shows the crash rates of AV in NDE and NADE, respectively. The crash rate in NDE is presented as the black line in Fig.~\ref{fig:accident_rate_NDE_NADE}, with the bottom $x$-axis as its number of tests. The blue line in Fig.~\ref{fig:accident_rate_NDE_NADE} represents the crash rate in NADE, and the top $x$-axis is the number of tests. The light shadow gives the 90\% confidence interval. It can be seen that the crash rates in NDE and NADE converge to the same value, while NADE requires a much smaller number of tests.
To measure the estimation precision of the crash rate, the relative half-width (RHW) \cite{zhao2016accelerated} is adopted as the metric. The threshold of RHW is set to 0.3. To reach this threshold, NADE requires 6.76\,$\times$\,10\textsuperscript{6} number of tests, while NDE requires 1.21\,$\times$\,10\textsuperscript{8} number of tests, as shown in Fig.~\ref{fig:relative_half_width_NDE_NADE}. It can be found that NADE can accelerate the evaluation by about 17.90 times compared with NDE. We note that the acceleration ratio is smaller than that in \cite{feng2021intelligent}, because combinations of multiple various SMs are applied in this paper, which improves the robustness yet decreases the efficiency. The goal of the adaptive evaluation is to improve the efficiency while keeping the robustness.

\begin{figure*}[!t]
  \centering
  \includegraphics[width=18cm]{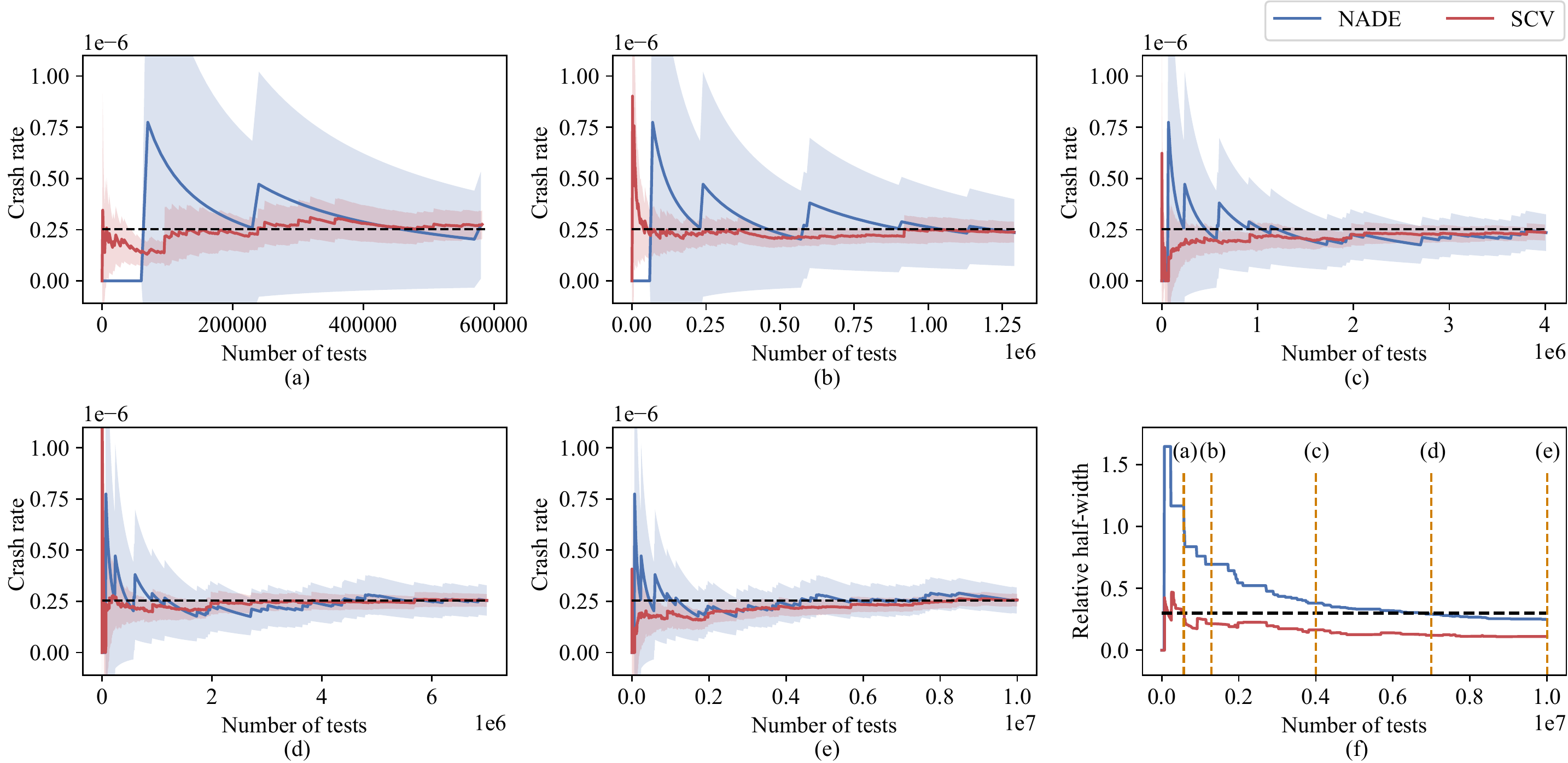}
  \caption{Crash rate of AV using NADE and SCV for (a) $n=5.92\times10^5$, (b) $n=1.29\times10^6$, (c) $n=4\times10^6$, (d) $n=7\times10^6$ and (e) $n=1\times10^7$, where $n$ is the total number of tests and the dashed line is the crash rate estimated by NDE; (f) RHW of AV evaluation using NADE and SCV, where the dashed line in black represents the RHW threshold (0.3) and 5 dashed lines in orange correspond to (a)-(e).}
  \label{fig:accident_rate_rhw_NADE_CV}
\end{figure*}

\begin{figure*}[!t]
  \centering
  \includegraphics[width=16cm]{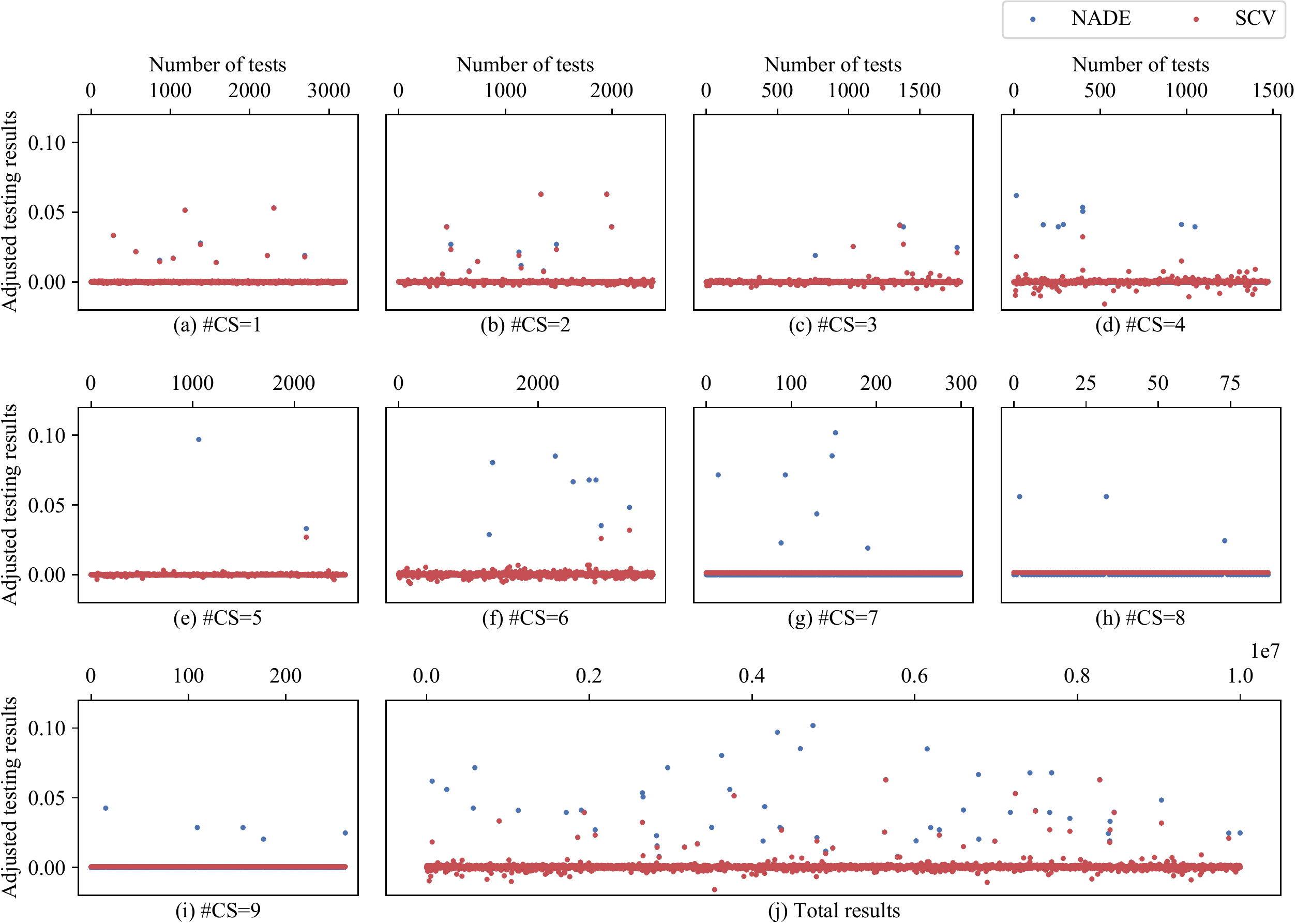}
  \caption{Adjusted testing results by NADE and SCV for (a)-(i) the number of control steps (\#CS) from 1 to 9 and (j) total 10\textsuperscript{7} testing results.}
  \label{fig:accident_NADE_ATSCV}
\end{figure*}

To investigate the performance of the SCV method, the accuracy and efficiency of AV evaluation in NADE with and without SCV are compared. It can be seen in Fig.~\ref{fig:accident_rate_rhw_NADE_CV} (a)-(e) that the crash rates of NADE and SCV converge to the same value for different number of tests. Fig.~\ref{fig:accident_rate_rhw_NADE_CV} (f) shows that the required numbers of tests of NADE and SCV for reaching the RHW threshold are 6.76\,$\times$\,10\textsuperscript{6} and 5.92\,$\times$\,10\textsuperscript{5}, respectively, resulting in a further acceleration ratio of 11.42. 
The weighted testing results before and after being adjusted by SCV with different number of control steps are compared in Fig.~\ref{fig:accident_NADE_ATSCV} (a)-(i), and Fig.~\ref{fig:accident_NADE_ATSCV} (j) shows the total 10\textsuperscript{7} adjusted testing results. It can be seen that the SCV method is able to adjust the testing results into a much narrower interval, especially for relatively large number of control steps (e.g., $l\geqslant4$), resulting in a considerable reduction of the estimation variance.

\begin{figure}[!t]
  \centering
  \includegraphics[width=8cm]{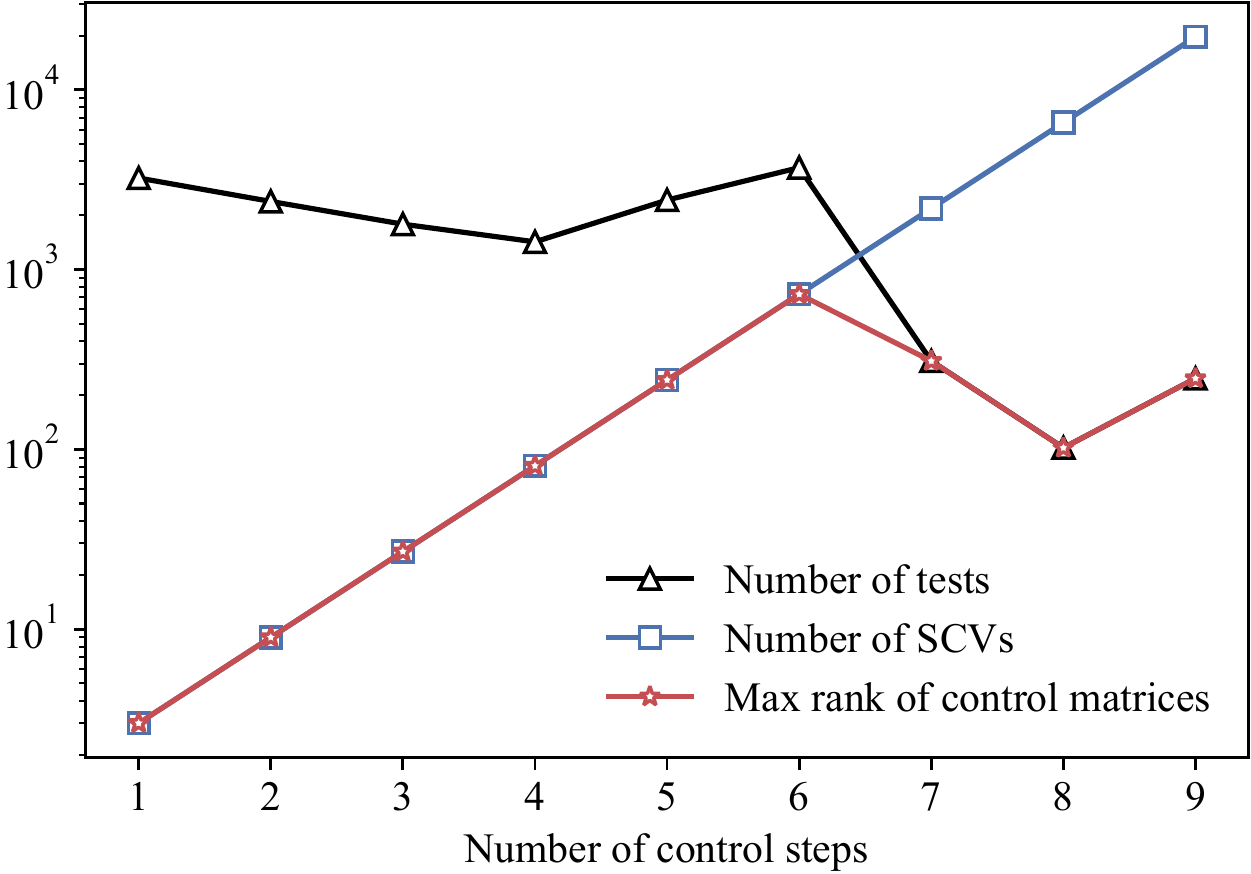}
  \caption{Number of tests, number of SCV and maximum rank of control matrices for different number of control steps.}
  \label{fig:num_tests_vs_max_rank}
\end{figure}

The detailed regression processes of the SCV method are also investigated. Fig.~\ref{fig:num_tests_vs_max_rank} shows the number of tests, the number of SCV and the maximum rank of the control matrices for the number of control steps $l=1,\dots,9$, respectively. Note that for $l\geqslant10$, we only use the first 9 control steps to construct the SCV. It can be seen that the maximum number of tests appears at $l=6$ and then the number of tests decreases to a relatively low level. As shown in Eq.~(\ref{eq:min_num_test_dimen}), the maximum rank of the control matrices is the minimum value between the number of tests and the number of SCVs, and hence will not grow exponentially with the number of control steps, although the number of SCVs will do. Therefore, the SVD of control matrices is always tractable in each stratum and the optimal control parameters can be found to minimize the estimation variance.

Since the scenario generation processes are stochastic, the testing and evaluation results are usually not the same in different experiments. Therefore, to find the average performances, we shuffle the testing results 200 times to bootstrap them and obtain the frequency distributions of the required number of tests (RNoT) in NDE and NADE. The average RNoT of NDE and NADE are 1.20\,$\times$\,10\textsuperscript{8} and 8.71\,$\times$\,10\textsuperscript{6}, respectively. Therefore, the average acceleration ratio (AAR) of NADE with respect to NDE is 13.78. The testing results of SCV are also bootstrapped by 200 times. For cases with maximum RHW below 0.3, we use the RNoT when the maximum RHW is reached. The average RNoT of SCV is 1.29\,$\times$\,10\textsuperscript{6}, resulting in an AAR of 6.76 times compared with NADE.

\subsection{Generalizability Analysis}

\begin{table}[!t]
  \centering
  \renewcommand{\arraystretch}{1.3}
  \caption{AARs of SCV where AV admits IDMs with different $\alpha$ values, and the rightmost column corresponds to the VT-IDM.}
  \label{tab:average_accelerated_rate}
  \begin{tabular}{c|cccccc||c}
    \hline
    $\alpha$ & 0.5&   1.0&  1.5&   2.0&  2.5&  3.0 & VT-IDM \\
    AAR & 11.52& 9.02& 7.87& 6.76& 7.73& 10.90 & 7.30 \\
    \hline\hline
    $\alpha$ & 3.5&   4.0&   4.5& 5.0&  5.5&  6.0 & \\
    AAR & 13.44& 11.95& 11.12& 10.61& 10.45& 10.05 & \\
    \hline 
  \end{tabular}
\end{table}

In the above experiments, we have set the AV model the same as SM-I, i.e., they are both IDMs with same parameters. To investigate the generalizability of the SCV method for different AV models, the IDMs with a series of parameters $\alpha=0.5,1.0,\dots,6.0$ are chosen as AV models. The AARs of SCV compared with NADE are shown in Table \ref{tab:average_accelerated_rate}. The testing results of all AV models are shuffled 200 times to obtain the AARs. It can be seen that the minimum AAR appears at $\alpha=2.0$, where the AV model is the same as SM-I, while the maximum AAR appears at $\alpha=3.5$. The mean AAR for different AV models is 10.12. Therefore, the SCV method can further accelerate the evaluation process by about one order of magnitude for various types of AV models. Moreover, the AARs of SCV with AV models different from SM-I are always greater than that of AV model the same as SM-I. The reason is that although using AV models different from SM-I will do harm to both the estimation efficiency of NADE and SCV, the damage to NADE is more than to SCV.

In addition, we also select the calibrated IDM in \cite{sangster2013application} (denoted as VT-IDM) as the AV model to further validate the generalization performance of the SCV method. The testing results shuffled 200 times give an AAR of 7.30 for SCV compared with NADE, which is shown at the rightmost column in Table \ref{tab:average_accelerated_rate}. Therefore, the SCV method can also increase the evaluation efficiency considerably for AV model with completely different calibrated parameters. This is not a surprising result because the only requirement for the SCV method to work is that the SMs and the AV model have some correlation, and more correlation contributes to more variance reduction. Although the VT-IDM and IDM have totally different parameters, they are still correlated to some extent.

\section{Conclusion}
\label{sec:conclusion}

In this paper, we propose an adaptive safety evaluation framework for CAVs in high-dimensional scenarios with a newly developed sparse control variates (SCV) method. To address the CoD, the SCV are constructed by only considering the sparse and critical variables of testing scenarios and stratified into strata accordingly. By optimizing the SCV leveraging the testing results within each stratum, the estimation variance is significantly reduced for different CAVs adaptively, accelerating the evaluation process. The accuracy, efficiency and optimality of the proposed method are verified and validated by both theoretical analysis and empirical studies. Comparing with the evaluation efficiency in NDE and NADE, our method is always more efficient particularly for CAVs that are different from SMs. It has been noted that adaptive testing scenario generation and adaptive testing result evaluation are two complementary approaches for adaptive testing and evaluation of CAVs. How to develop the former in high-dimensional scenarios deserves further investigation.

\bibliographystyle{IEEEtran}
\bibliography{IEEEabrv,reference.bib}

\end{document}